\documentstyle[12pt]{article}

\def\be{\begin{equation}}
\def\ee{\end{equation}}
\def\b{\beta}
\def\d{\delta}
\def\i{\infty}
\def\k{\kappa}
\def\l{\lambda}
\def\o{\omega}
\def\s{\sigma}
\def\s{\sigma}
\def\e{\epsilon}
\def\f{\varphi}
\def\t{\theta}
\def\r{\rho}
\def\p{\partial}

\begin{document}
\vspace{1.5in}
\begin{center}

{\large \bf  Nonlinear Hall effect in an AC electric field and related
phenomena.}

\vspace{0.2in}

{ A.A.~Ovchinnikov }
\vspace{0.2in}

{\it  Max Planck Institute for Physics of Complex Systems, Dresden.}

\vspace{0.2in}
\end{center}

\begin{abstract}

It has been shown that in metals and semiconductors the joint action of
permanent magnetic and AC electric fields leads to arising of DC surface
electric current. The physical reason for such a current is due to
essentially non-linear dynamics of electronic gas in the surface layer. In
adiabatic limit the analytical expression for the current are
given.  Two cases have been considered: a surface current inside
metal at the metal-insulator boundary and a current in an injection layer on
metal- semiconductor boundary. General equations of electronic
gas dynamics have been solved numerically. The surface current is linear with
respect to magnetic field and depends quadratically on an amplitude of
external AC electric field. In inhomogeneous media with non-zero gradient of
conductivity there can be the bulk current. Corresponding phenomenological
expression is obtained.The conditions for spontaneous generation of magnetic
field have been found.

\end{abstract}

\vspace{0.12in}


\vspace{0.12in}

The directed transport of particles under action of AC driving force is now
well established phenomenon in various non-linear systems. Most detailed this
transport has been studied for one-dimensional motion of particle in the
space-periodic potential $U(x)$ and a time-periodic force having general form
$F(t)=F_0 \cos{\o t}+F_1 \cos{(2\o t+\t)}$. (See review \cite{1} and recent
works \cite{2,3,4}.) Main physical reason for the directed motion of the
particles is connected with the so called latent asymmetry of driving force
$F(t)$. Though the time average of $F(t)$ is zero the time average
$\overline {F^3(t)}$ doesn't vanish and depends on $\t$. Non-linearity of
the system with potential $U(x)$ leads to mixing of harmonics such a way that
average velocity is not zero and directed along $\overline {F^3(t)}$   .The
necessary symmetry conditions for $U(x)$ and $F(t)$ are considered in
\cite{4}.

The aim of this paper is to show that the phenomenon can lead to various
experimentally observable physical consequences in solid state physics and
physical chemistry. In particular, the non-linear Hall effect could take
place in surface layer of metals and semiconductors when harmonic AC
electric field applied instead of permanent electric field used in
conventional Hall effect experiments. The set up and system of coordinate
used are shown in Fig.1.

Firstly we calculate the non-linear Hall (NLH) current in skin layer of
metals on the metal-insulator  boundary. Then we give a consideration of NLH
effect in an injection layer of semiconductor at the metal-semiconductor
boundary.And finally, we show that even neutral nano-size particles
can move along a surface perpendicularly to both the magnetic and AC electric
fields. We estimate also corresponding driving forces.

\vspace{0.02in}
     {\bf 1. Skin layer NLH effect.  }.
\vspace{0.02in}

Consider dynamics of electronic gas in skin layer of metal under
an influence of AC electric field directed perpendicularly to surface of the
metal and in the magnetic field. Since the width of a  skin layer and all
other characteristic lengths greater than atomic distances one can readily
use hydrodynamics approximation for description of system.

In the time-periodic steady state regime there are three variables we need to
calculate: density of electrons $\r (x,t)$ and two components of velocity
$v_x=v(x,t)$ and $v_y=u(x,t)$. All variables are functions of only one
coordinate $x$ and a periodic  function of $t$ with frequency $\o$ of the
external field $E_0(t)$. The equations for them looks as following
\be
\r\Biggl({{\p v} \over {\p t}}+v{{\p v} \over {\p x}}\Biggr)=
-{{\p p} \over {\p \r}}{{\p \r} \over {\p x}}
-{{8\pi e^2} \over {m}} q(x)-{{e} \over {m\mu}}\r v
-{e\over m}E_0(t)\r-{{He} \over {mc}}\r u
+ \eta {{\p ^2 v}\over {\p x^2}}
\label{1}
\ee
\be
\r\Biggl({{\p u} \over {\p t}}+v{{\p u} \over {\p x}}\Biggr)=
{{He} \over {mc}}\r v -{{e} \over {m\mu}}\r u +
\eta {{\p ^2 u}\over {\p x^2}}
\label{2}
\ee
\be
{{\p \r} \over {\p t}} = -{{\p } \over {\p x}} (v\r)
\label{3}
\ee
Here $p$ is a pressure of electronic gas in metal, $\p p/\p\r=c_e^2$ is a
square of a sound velocity of electronic gas, $\mu$ is a mobility of
electron in metal, c is the light velocity, m is effective mass of electrons,
$\eta$ is a viscosity coefficient of an electron liquid, $H$ and
$E_0(t)=E_0\cos{\o t}$ are external magnetic and AC electric fields,
respectively. A quantity

\be
q(x,t)=\int_0^x (\r(x',t)-\r_0)dx'
\label{4}
\ee
is a surface density of charge (together with positive charge of
lattice) accumulated in a layer between $x=0$ and $x$.

The boundary conditions for this system are quite simple. Both velocities
$v,u$ are equal to zero at the edge of metal ($x=0$).

Taking the limit $x\to \infty$ we come to three equations for asymptotic
values $q_{\i}=q(\i,t)$, $u_{\i}=u(\i,t)$,
$v_{\i}=v(\i,t)$
\be
{{d}\over dt}q_{\i}=-\r_0v_\i
\label{5}
\ee
\be
{{d}\over dt}v_{\i} = -v_\i\o_{\mu}+{e\over m}E_0(t)-2\b q_\i(t)
\label{6}
\ee
\be
{{d}\over dt}u_{\i}=v_\i\o_L-u_\i\o_{\mu}
\label{7}
\ee
Here $\o_{\mu} =e/m\mu$ and $\o_L=eH/mc$ is a Larmor frequency.

Periodic solutions of these linear equations give the boundary conditions at
$x\to \i$. And $q(0,t)=0$ by definition.
Note that we omitted magnetic term in a r.s. of equation (6) because we are
interested only in linear with respect to $H$ contribution to the Hall
current. Since one needs only the periodic solution of (1-3) the initial
conditions are arbitrary.

The system of equations (1-3) has been analyzed both by perturbation theory
with respect to external electric field and numerically using standard NAG
routines.

At weak external AC electric and magnetic fields the total surface NLH current
$I_H$ could be expressed as following
\be
I_H=c^2_e n m \mu \Biggl( { {\mu E_0} \over {c_e} }\Biggr)^2
\Biggl({{\mu H}\over {c}}\Biggr) \k (\o,\b )
\label{8}
\ee
where
\be
c_e=\Biggl({1\over m}{\p p\over \p\r}\Biggr)^{1/2}
\label{9}
\ee
is the sound velocity of electronic gas in metal.
At low temperature $c_e$ is proportional Fermi velocity and doesn't depend on
temperature $T$. For Boltzmann electronic gas $c_e\sim \sqrt{T}$.  We prefer
here to express $I_H$ via a mobility of electrons in a skin
layer. For metal it might be more convenient to express it via the bulk
conductivity $\s$ of the material using a relation $\s=e\mu n$. A
dimensionless coefficient $\k$ is a function of two dimensionless parameters
$\o/\o_{\mu}$ and $\b=4\pi n \mu^2 m$. At low frequency of external AC
electric field, i.e. at $\o$ much less than the surface plasmon frequency,
$\k\sim(\o/\o_{mu} )^2$. In opposite limit $\k$ decays as
$\sim(\o_{\mu} /\o)^4$.  Main temperature dependence of $I_H$ comes from a
temperature dependence of mobility of electrons in the skin layer of the
material. Using reasonable parameters for $E_0$ ($\sim100~ V/cm$), $H$
($\sim 1000 Gauss$), $\mu$ and  $c_e$ we see that total surface NLH
current could reach observable value of a few $\mu A/cm$.
The typical dependence of $\k(\o )$ is shown on fig.2. Note that $\k<0$ 
which  
means that the corresponding current is not diamagnetic one. The physical
consequence of this fact will be discussed below. The profile of density of
NLH current as function of distance from the edge of a metal is shown on
Fig.~3. The space oscillations show that the viscosity of
electronic gas could play a significant role in the effect. The corresponding
discussion will be given below.

\vspace{0.02in}
     {\bf 2. NLH current in an injection layer}.
\vspace{0.02in}

The essentially different physical situation takes place in an injection
layer at the boundary metal-semiconductor. In the thermal  equilibrium the
free energy (per unit area) of electron gas  can be written as following
\be
F=\int_0^\i \e (\r )dx
+2\pi e^2\int_0^\i \int_0^\i \r (x)|x-x'| \r (x')dxdx'
+4\pi e^2\s_0\int_0^\i x\r (x)dx
\label{10}
\ee
where
\be
\s_0=\int_0^\i \r (x) dx
\label{11}
\ee
is a surface density of the injected electrons.

The equality of chemical potentials of metal and semiconductor on the
boundary ($x=0$)   gives a condition for definition of $\s_0$ and $\r_0=\r
(0)$
\be
W={{\p \e } \over {\p\r}}\Biggl|_{x=0}+4\pi e^2\int_0^\i \r (x) x dx
\label{12}
\ee
$W$ being a difference of the work functions of metal and semiconductor.
Minimisation of F gives the following result for equilibrium density of
injected electrons (in a case of Boltzmann electron gas )
\be
\r_{eq}(x)=\r_0 \cosh^{-2}(x/x_d)
\label{13}
\ee
where $x_d$ is a Debay length of a double layer
\be
x_d=(kT/4\pi e^2\r_0)^{1/2}
\label{14}
\ee
Under action of time-periodic external electric field $E_0\sin{\o t}$ the
non-linear oscillations of the density $\r (x,t)$ takes place. A system of
equations describing these oscillations looks very much like that of
the metal-vacuum interface equations (1-4), but the background positive
charge $\r_0$ in Eq.(4) has to be taken zero.

Another important difference between the metal-insulator and
metal-semiconductor cases is connected with boundary conditions for
velocities $v(x,t)$ and $u(x,t)$ at $x=0$. There are two limit cases.
In the first one, the exchange of charges over the surface is quite slow
which means that total charge in the semiconductor is conserving under action
of the AC electric field and $v(0,t)=0$, $u(0,t)=0$ at the boundary. In this
case if we neglect the viscosity a total NLH current vanishes. However,
taking into account small viscosity leads to the following non-zero surface
NLH current
\be
I_H = {{e c_e^2} \over {\o_{\mu}m}} \eta
\Biggl({{E_0\mu} \over {c_e}}\Biggr)^2
\Biggl({{H\mu} \over c}\Biggr)
\Biggl({{\o_{\mu}} \over {c_e^2}}\Biggr)\k(\b,\o/\o_{\mu})
\label{15}
\ee
$$
\b={{e^2\r_0} / {m\o_{\mu}^2}},~~~
\r_0 ={1\over 2}\r_T exp\Bigl( W/kT \Bigr)
$$
So, the surface NLH current in this case depends linearly on viscosity of an
electronic gas in a double layer. The dimensionless coefficient
$\k(\b,\o/\o_{\mu})$ as function of frequency of external electric field
looks very much like that of a metal. It is proportional $\o^2$ at small 
$\o$ and decays as $1/\o^4$ in a high frequency regime.

If the rate of exchange of electron through the surface metal-semiconductor
is very large the density of electron at the surface is a constant
which equals to
its equilibrium value $\r_0$. The analytical treatment in adiabatic
approximation ($\o$ is much less than plasmon frequency of the double layer)
and numerical calculations give the quantitative expression for
$\k(\o/\o_\mu )$. The details of these calculations will be given elsewhere.

\vspace{0.02in}
     {\bf 3. NLH effect phenomenology and generation of  magnetic field }.
\vspace{0.02in}

Thus, at simultaneous application of magnetic and AC electric field on the
boundary separating two media the surface NLH current is arising. The effect
is taking place also in a bulk of material if there is a nonzero gradient of
conductivity tensor $\s_{ij}$. The invariant expression of the NLH current
density could be written as
\be
j_i=\l (\o ){{\p \s_{ij}} \over {\p x_k}}{\dot E}_k {\dot E}_l
\e_{jlm}H_m
\label{16}
\ee
where ${\dot E}_k$ is a time derivative of an external electric field and $H$
is magnetic field.  In a case of isotropic medium $\s_{ij}=\d_{ij}\s$ the
density of current is
\be
{\bf j}=\l ({\bf \nabla}\s {\dot {\bf E}}) [{\dot {\bf E}}\times {\bf H}]
\label{17}
\ee
Here $\l$ is a phenomenological coefficient depending on frequency and on the
material parameters.

Let us stress once again that this current is not necessary diamagnetic one.
The direction of current depends on a sign of the conductivity gradient.
So, the magnetic field produced by this current could increase the external
magnetic field giving a mechanism of spontaneous generation of magnetic field
(even in absence of external magnetic field).

In order to give an estimation of generated magnetic field let us consider
infinite metalic cylinder as it is shown on Fig.1 and assume that the
conductivity of cylinder is decreasing along the radius ($d\s /dr<0$). The
external AC electric field also directed along the radius. In cylindrical
coordinate system there is only angular component of the current density
$\overline {j_{\f}}(r)$ averaged over time. According to (17) it is
\be
{\overline {j_\f}}(r)=\l {{d\s}\over {dr}}{\overline {{\dot E}^2}}H
\label{18}
\ee
The magnetic field $H$ in this expression is directed along $z$ axis and can
be a function of radius $H(r)$. The additional magnetic field $H_{ad}(r)$
induced by this current has to be determined from magnetostatic equation
\be
curl~{\bf H}_{ad}(r)={{4\pi}\over c} {\bf j}(r)
\label{19}
\ee
with the general solution
\be
H_{ad}(r)=-{{4\pi}\over c}\int_{r_1}^r j(r') dr'
\label{20}
\ee
>From (20) it is clearly seen that if $d\s /dr<0$ the induced magnetic field
$H_{ad}(r)$ increases an initial magnetic field H(r) giving possibility of
generation of magnetic field. Thus, at some condition the NLH
current can be observable even without external magnetic field.   Moreover
the AC electric field could be substituted by some acoustic vibrations of
electronic gas, which leads to, so to say, the NLH current without both
magnetic and electric fields.

Coming back to a surface situation we would like to note that a directed
current of particles along the surface is possible even if we deal with
neutral particles. In this case the nonlinear vibration of positive and
negative charges inside the neutral particle could be quite different giving
the Lorentz forces of different values. As a result, it creates driving force
directed perpendicularly to magnetic field. The quantative consideration of
this effect will be given elsewhere.

The main obstacle for observation of the NLH effect in AC electric field
seems to be connected with an absorption of ultrahigh frequency
electromagnetic field into material. The time-periodic components of Hall
current which is linear with respect to external electric field is greater
than a permanent component of the NLH current and may lead to heating up of
the sample. On the other hand, at low (or ultralow) temperature the
absorption and warming up could be made small enough for such observation.

Author thanks M.Ya.Ovchinnikova for help in numerical calculations and
S.Flach for numerous discussions on related problems.

\vspace{0.2in}

%\newpage

\vspace{0.2in}

%\newpage
{\bf Captions to Figures}
\vspace {0.15in}

Fig.~1.
The principle set up for observation of NLH effect. Electric
field $E_0$ and magnetic field $H$ are directed along $x$ and $z$ axes,
correspondingly. In case of large radius the problem of current near the
interface reduces to one-dimensional problem discussed in text.

Fig.~2.
Total surface NLH current (in unit $[e\r_0c_e^2/\o_{\mu}]$) as function
of frequency $\o/\o_{\mu}$ of external electric field at $\b=2.0$, $\eta=0$.

Fig.~3.
The current density profile $j(x)$  (in unit $[ec_e\r_0]$) as function of
distance x (in unit $[c_e/\o_{\mu}]$) at $\o/\o_{mu}=0.6$ and $\b=2.0$,
$\eta=0$.

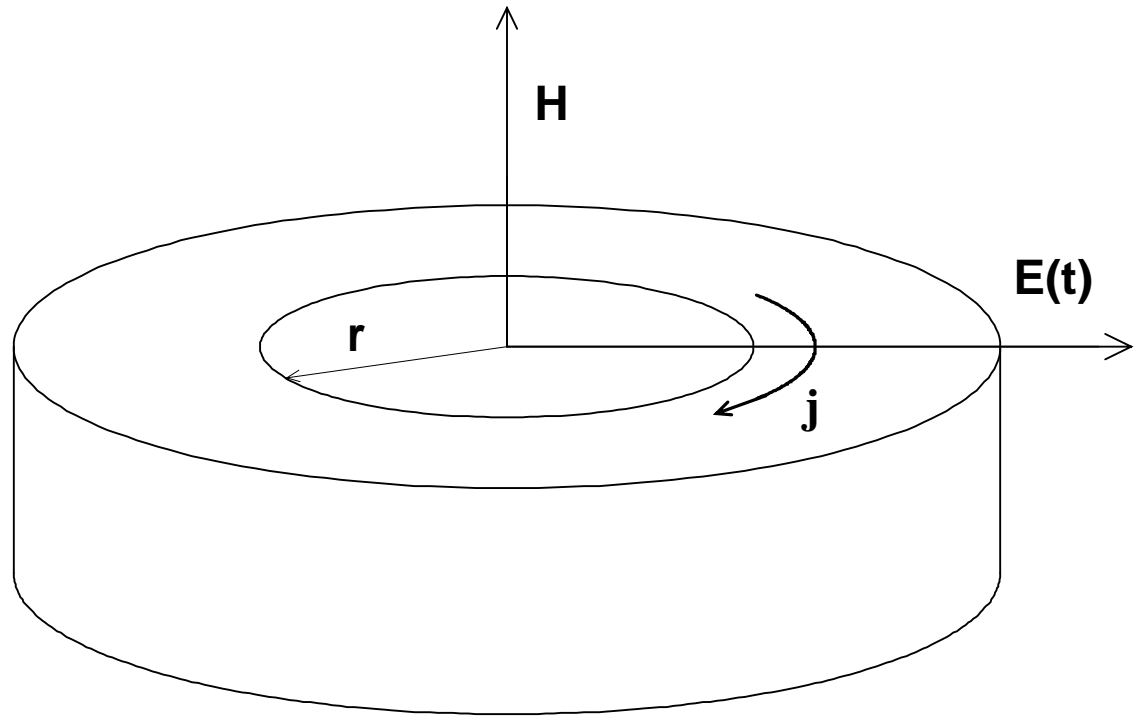
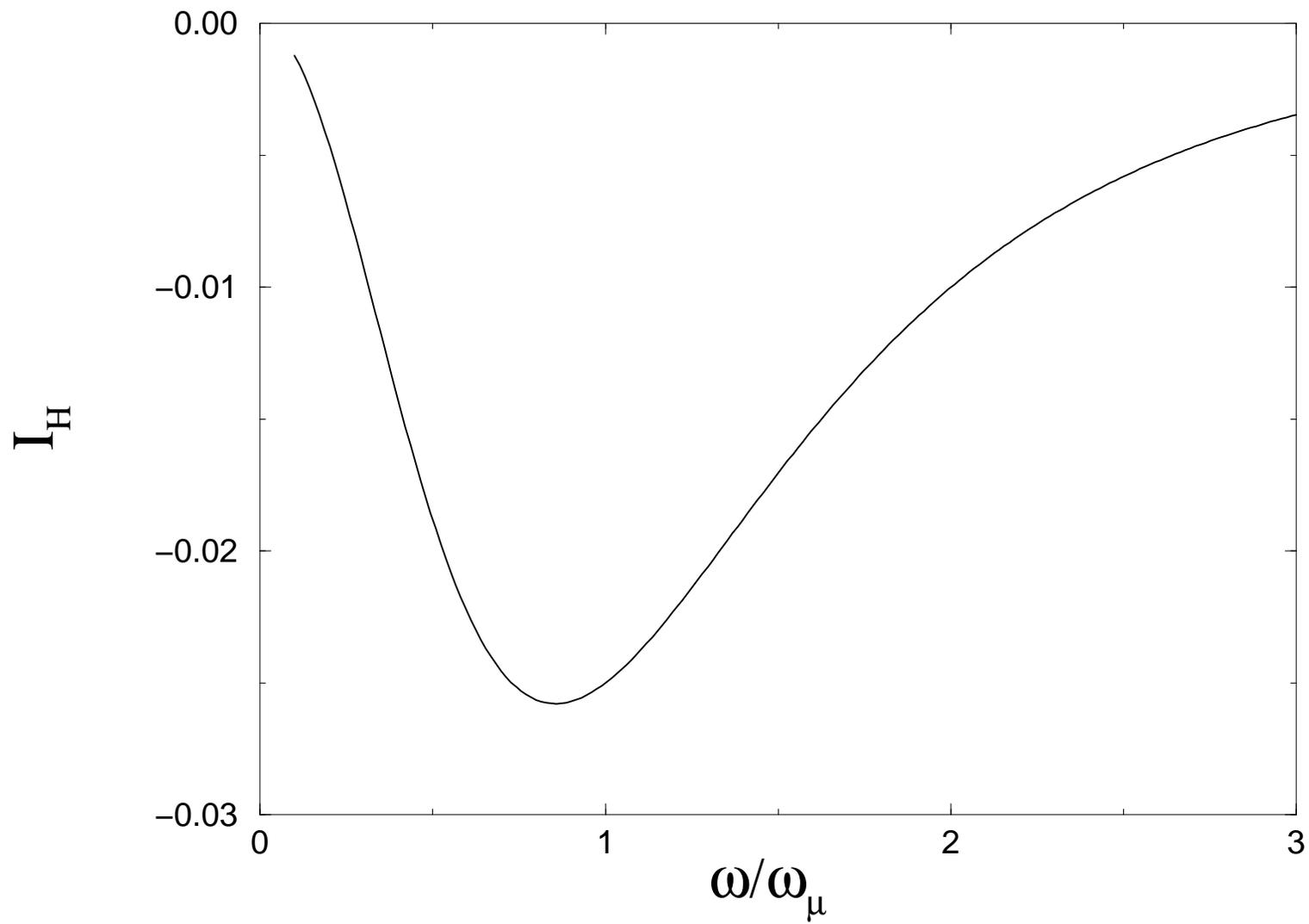
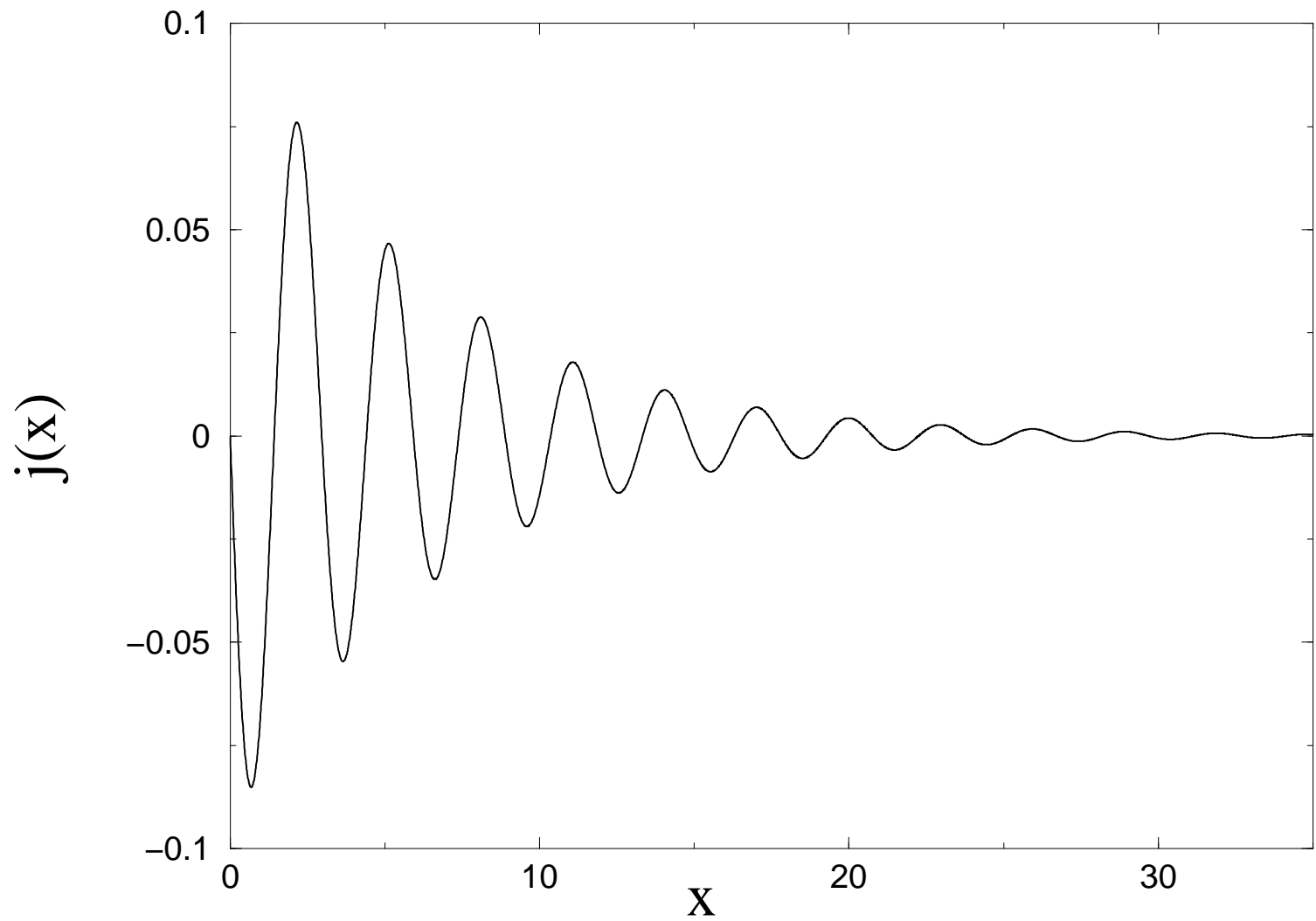
\end{document}